\documentclass[%
  reprint,
 superscriptaddress,
  amsmath,amssymb,
  aps,
  pre,
  longbibliography,
 ]{revtex4-2}

 \usepackage{graphicx}% Include figure files
 \usepackage{dcolumn}% Align table columns on decimal point
 \usepackage{bm}% bold math
 \usepackage{float}
 \usepackage{color}

 \newcommand{\eps}{\varepsilon}

\begin{document}

 \preprint{APS/123-QED}

 \title{
%Lengthscale Competition in Tunable Polymeric Interactions: From Microscopic Properties to Mesoscopic Phases
 %or\\
Tuning polymer architecture for quasicrystal self-assembly
% Designing emergent phases via tunable micellar interactions\\
% or\\
 %Soft matter design across scales: connecting micellar interactions to emergent phases
 }% Force line breaks with \\

 \author{D. J. Ratliff}%elliot\email{d.j.ratliff@northumbria.ac.uk}
 \affiliation{Department of Mathematics, Physics and Electrical Engineering, Northumbria University, Newcastle upon Tyne, NE1 8ST, United Kingdom}

 \author{A. Scacchi}
 \affiliation{Department of Mechanical and Materials Engineering,
University of Turku, Vesilinnantie 5, FI-20014 Turku, Finland}

 \author{P. Subramanian}%\email{priya.subramanian@auckland.ac.nz}
 \affiliation{Department of Mathematics, University of Auckland, 38 Princes Street, Auckland 1010, New Zealand}

 \author{A. J. Archer}
 \affiliation{Interdisciplinary Centre for Mathematical Modelling and Department of Mathematical Sciences, Loughborough University, Loughborough, Leicestershire, LE11 3TU, United Kingdom}

 \author{A. M. Rucklidge}
 \affiliation{School of Mathematics, University of Leeds, Leeds LS2 9JT, United Kingdom}

 \date{\today}% It is always \today, today,

 \begin{abstract}
 Using computer simulations and theory, we investigate the ultrasoft interactions between dendrimers formed of a central polymer connected by stiff linkers to a corona of flexible polymers, forming `pompoms' at the ends of the linkers. %[micelle-like aggregates formed from diblock copolymers].
 We show that the resulting coarse-grained interaction potential between pairs of dendrimers exhibits tunable lengthscale competition based on properties of the core and corona polymers. %each polymer block.
 We present a simple model for this pair potential, which we confirm using accelerated Monte Carlo methods. We then demonstrate the connection between dendrimer structure and mesoscopic phases by presenting parameter choices that result in stable dodecagonal quasicrystals, and show that the size of the region in the phase diagram where quasicrystals are stable can be controlled by tuning details of the polymer architecture alone. 
These results pave the way for experimental realization of soft matter quasicrystals by identifying what overall molecular architecture leads to their stability.
 %\textcolor{cyan}{These results pave the way towards improving the experimental realization of soft matter quasicrystals by highlighting how the molecular chemistry influences their stability.} 
 \end{abstract}

 %Use showkeys class option if keyword
 \maketitle

Fundamental understanding of how soft matter can self-assemble to form aperiodic quasicrystals (QCs) comes from studies of model systems of interacting soft penetrable particles \cite{bdl11, ark13, bel14, ark15, Savitz2018, Ratliff2019, Subramanian2021}.
These have pair potentials $V(r)$ that are purely repulsive and finite for all particle separations distances~$r$.
Potentials of this general form arise as mesoscopic effective interaction potentials between the centers of mass of polymeric macromolecules \cite{l01}.
These models exhibit phase diagrams containing not only QCs, but also a variety of other structures, such as periodic crystals and lamellar phases.
However, there remains a natural and fundamental question that needs to be answered: how might these QC-forming potentials~$V(r)$ be connected to the internal architecture of the polymers and also to the lengthscales of the emergent crystal phases? In other words, what are the general features of the polymers that, when coarse-grained, lead to effective potentials that may give rise to QCs?

Coarse-graining (in the statistical mechanics sense) corresponds to integrating over some of the degrees of freedom in the system and generally results in a much simpler description in terms of the remaining degrees of freedom, i.e., the centers of mass of the polymers and the effective potential~$V(r)$ between these.
The resulting effective~$V(r)$, together with a simplified dynamics, can then be input into simulation or theory to determine the overall phase behavior.
Suitable approaches include molecular dynamics simulations and density functional theory (DFT)~\cite{hansen_mcdonald, evans_79, evans_92, lutsko2010recent}. 
For linear polymers, and branched polymers such as dendrimers, the form of the coarse-grained~$V(r)$ between the centers of mass of the molecules are fairly well understood \cite{l01}, as is the collective phase behavior.
Understanding why crystallization occurs comes from analysis in Fourier space and consideration of~$\widehat{V}(k)$, the Fourier transform of~$V(r)$. 
When $\widehat{V}$ possesses a negative component with a minimum at $k_1$, a Turing-like instability is possible, which results in the emergence of periodic (or aperiodic) structures having a lattice constant, or lengthscale, of $2\pi/k_1$~\cite{lmg07,mgk07,mlk08, archer2014solidification, archer2016generation, scacchi2018flow}.
It is now understood that having two (or more) competing lengthscales, with a specific ratio $k_1/k_2$, can encourage the formation of QCs~\cite{Lifshitz2007,bdl11,Subramanian2016}.
Having more than one lengthscale arises naturally with more complex molecular architectures, including diblock copolymers, star-polymers, dendrimers and surfactants, which include some of the soft matter systems where QCs arise~\cite{zul04, hdt07, fez11, ge11, ikg11, zb12, llb14, glb16,  yhm16, hyw18, llr20, jbm21, mlj21, hyw18, Joseph2023, Joseph2025}.
Of these systems, the candidates we focus on here (because, as we show, they have effective potentials of the requisite form) are star-polymers and dendrimers~\citep{l01, gotze2004tunable, mlk08} (sometimes referred to broadly as `soft colloids'~\cite{vlassopoulos2014tunable}) and also certain micelles formed from diblock copolymers~\cite{pah06, cph09, scacchi2021self,chang2020impact}.
In particular, this letter provides a theory and route-map for how to design polymeric macromolecules with the architecture tuned so that they self-assemble to form QCs, a phase that is hard to manufacture and stabilize in experimental settings.
We also outline how smart Monte Carlo (MC) approaches can be taken advantage of in order to optimize the coarse-graining stages of the problem.
 
\begin{figure}[t!]
 \centering
 \includegraphics[width=0.9\columnwidth]{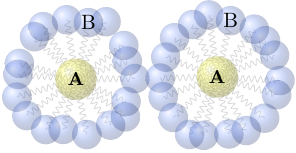}
 \caption{A sketch of a pair of $N = 18$ arm molecules, each with the yellow core A particle tethered to $N$ blue B particles. These have radii $R_{\rm AA}$ and $R_{\rm BB}$, respectively; c.f.\ Eq.~\eqref{pair-pots}. The potentials in Eq.~(\ref{tether-pot}), tethering each A to the $N$ surrounding B particles, are represented as gray springs.}
 \label{fig:micelle}
 \end{figure}

 We concern ourselves with a simplified polymeric molecule that has a structure illustrated in the coarse-grained picture shown in Fig.~\ref{fig:micelle}.
 It is composed of a core part (yellow sphere, which we denote A), that may in some cases be point-like (as in star polymers), from which $N$ `stiff' arms originate. These are illustrated as the gray springs in Fig.~\ref{fig:micelle}. Here, `stiff' means polymer segments that have a length less than the persistence length \cite{l01, Cloizeaux1990}.
 These linker arms are then assumed to subsequently connect to more flexible polymers, which may be branched (as in the case of, e.g., dendrimers), or continue as long and coiled linear chains (such as star polymers).
 These flexible end polymer `pompoms' (denoted B) are illustrated as the blue spheres in Fig.~\ref{fig:micelle}, where the sphere radii correspond roughly to the radius of gyration of the B-polymers.
 We assume that the core of the molecule (A), as well as the terminal part of the arms (B), can all be modeled by soft repulsive interactions between their local centers of mass, with the following Gaussian form
 \begin{equation}\label{pair-pots}
 \phi_{ij}(r) = \eps_{ij}e^{-\frac{1}{2}\big(\frac{r}{R_{ij}}\big)^2}\,, \quad i,j={\rm A,B}\,,
 \end{equation}
 where $\eps_{ij}$ denotes the overlap penalty between polymer portions of type $i$ and $j$, $R_{ij}$ the interaction ranges, and $r$ the distance between their centers of mass.
 This potential is well suited to describe the effective interaction between both linear polymers and dendrimers in good solvents~\cite{l01, Louis20002522,Bolhuis20014296, gotze2004tunable,10.1063/1.2172596}. The core A and each terminal chain B are additionally tethered by the harmonic potential
 \begin{equation}\label{tether-pot}
 \phi_{\rm AB}^{\rm t}(r) = \frac{k_{\rm D}}{2}(r-\ell_0)^2\,,
 \end{equation}
 where $\ell_0$ is the natural extension of the tether, and $k_{\rm D}$ its stiffness.

A molecular architecture very similar to the one described above may also arise in the micellar structure of self-assembled amphiphilic diblock copolymers, as in~\cite{pah06}, where control over the pair interaction parameters in Eq.~(\ref{pair-pots}) ($\eps_{ij}$ and $R_{ij}$) can be achieved by varying the monomers in the different polymer blocks, the block lengths, as well as the solvent environment (change in solvophilicity and solvophobicity)~\cite{de1975collapse, de1978collapse, dautenhahn1994monte}.
Likewise, the tether potential can readily be modified by changing the diblock length ratio~\cite{capone2009entropic} or the miscibility properties of the two blocks~\cite{olaj1998lattice}. Note that a variety of more complex polymeric constructs can form dumbbell-shaped particles~\cite{kim2006nanorings, ge2007facile, zhang2007morphology,polymeropoulos201750th}. Similar micellar configurations as the one in~\cite{pah06} can also be obtained from a linear ABC tri-block terpolymer~\cite{fustin2005triblock}. In the case of dendrimers, this tunability is not solvent based but rather architectural, related to the generational number and length of the arms.
 
 \begin{figure*}[t!]
 \centering
 \includegraphics[width=0.95 \textwidth]{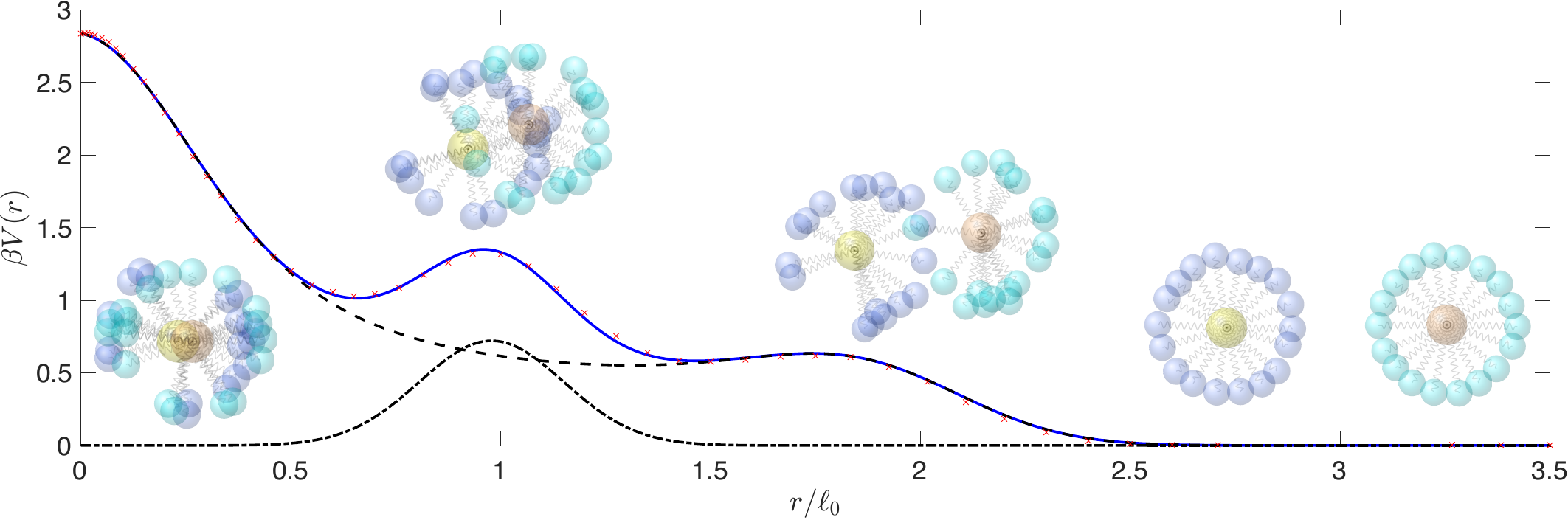}
 \caption{The effective coarse-grained pair potential $V(r)$ as a function of normalized distance between centers of mass $r/\ell_0$. Each molecule has $N=18$ arms, with parameters $(\beta\eps_{\rm AA},\beta \eps_{\rm AB},\beta\eps_{\rm BB}) = (0,5.4,5)$, $(R_{\rm AB}/\ell_0,R_{\rm BB}/\ell_0) =(0.06,0.045)$. The red points are the MC simulation results, while the blue solid line is the fitting using Eq.~(\ref{eqn:Coarse-grained-PP}), with parameters $(g_{\rm BB},g_{\rm AB},r_0/\ell_0,w/\ell_0) = (0.5002,0.8787,0.9902,0.2277)$. The black dot-dashed and dashed lines are the individual contributions from the second and third terms in Eq.~\eqref{eqn:Coarse-grained-PP}, respectively.
 Also displayed are snapshots from the MC simulations at separations $r/\ell_0 = 0$, 0.95, 1.9 and 3.25, highlighting the various interaction stages.}
 \label{fig:2}
 \end{figure*}

We obtain the effective interactions between pairs of macromolecules using both Brownian dynamics and Langevin MC computer simulations~\cite{rdf78,rt96,md08,sld20}; further details are given in the Supplementary Information (SI).
The presence of the tether potentials makes sampling difficult because they make large moves of the central A~particles unlikely.
Splitting the various separation distances between pairs of molecules into a sequence of `windows', and applying an umbrella bias potential to their centers of mass~\cite{mlk08}, facilitates fully sampling each window.
Thus, umbrella sampling allows us to accurately capture all of the features of $V(r)$, including in regions of high repulsion strength between the molecules.

Our aim is to establish the overall coarse-graining procedure of going from polymer architecture to phase diagrams. Thus, for simplicity, we assume that our system is two dimensional (2D), like the systems \cite{bdl11, ark13, bel14, ark15, Savitz2018, Ratliff2019, Subramanian2021}. However, our overall strategy can also be applied in three dimensions, at the cost of making the calculations more computationally expensive.
A typical result for the dimensionless effective dendrimer-dendrimer pair potential $\beta V(r)$, where $\beta=(k_BT)^{-1}$, $k_B$ is Boltzmann's constant and $T$ the temperature, obtained from MC simulations is displayed in Fig.~\ref{fig:2}, together with simulation snapshots of dendrimers at various pair-separation distances.
We also display in Fig.~\ref{fig:2} the potential obtained from an approximate (but rather accurate) coarse-graining procedure that is explained below.
We see that the molecules begin to interact at separations slightly larger than twice the natural extension of the tether,~$\ell_0$, as the B particles begin to push into each other.
As the separation distance~$r$ is decreased, $V(r)$~increases as more and more B's interact.
A peak in $V(r)$ arises at $r\approx\ell_0$, where the A and B particles are in close proximity with one another.
For small~$r$, as the molecules begin to completely overlap, a peak arises in~$V(r)$ at the origin.
Visually, the resulting $V(r)$ is structurally very similar to the exponentially damped degree~8 polynomial potential postulated by Lifshitz {\it et al.}~\cite{bdl11,bel14}, which is known to exhibit QCs in the phase diagram~\cite{rss19,Subramanian2021}.
Thus, this is an encouraging sign of the presence of competing, tunable lengthscales within systems involving such molecule-molecule interactions.

The next step is to explore the impact of the molecular architecture on the bulk (mesoscale) phase diagram. 
To accomplish this, we fit the MC data to an explicit functional form.
We could, e.g., consider the potential of Lifshitz {\it et al.}~\cite{bel14}, but the links between the molecular features and the functional form of the pair potential are lost in such fittings.
Moreover, it requires more than a degree~8 polynomial to correctly capture all the physical space and Fourier space details of the potential, which critically influence crystal formation~\cite{Ratliff2019}. 
Instead, to preserve the link between the molecular parameters and the functional form of the fit, we follow similar arguments to Likos {\it et al.}~\cite{lsl01,lrd02,l06}, in which the molecular architecture is modeled by considering only the one-body density distributions of the A and B particles within each macromolecule.
A mean-field treatment of the interactions between the particles is thus obtained, approximating two-particle distributions via products of one-particle distributions, as described in detail in the SI.
This leads to the following functional form for a fitted pair potential between molecules with centers located at ${\bf r}_{\rm C}^{(1)}$ and ${\bf r}_{\rm C}^{(2)}$:
\begin{align} 
&V({\bf r}_{\rm C}^{(1)}-{\bf r}_{\rm C}^{(2)}) = \phi_{\rm AA}(|{\bf r}_{\rm C}^{(1)}-{\bf r}_{\rm C}^{(2)}|) \nonumber\\
&\qquad{}+2g_{\rm AB}\int\phi_{\rm AB}(|{\bf r}_{\rm C}^{(2)}-{\bf r}|)\rho_{\rm B}^{(1)}({\bf r}) \,d{\bf r} \label{eqn:Coarse-grained-PP} \\
&\qquad{}+\frac{g_{\rm BB}}{2}\!\!\sum_{k,n=1}^2\!\int\!\!\!\!\int\phi_{\rm BB}(|{\bf r}-{\bf r}'|)\rho_B^{(n)}({\bf r})
    \rho_{\rm B}^{(k)}({\bf r}')
     \,d{\bf r}d{\bf r}' \nonumber
\end{align}
(see Eq.~(\ref{eq:int_energy_simp}) in the SI). 
Here, $\phi_{\rm AA}$, $\phi_{\rm AB}$ and $\phi_{\rm BB}$ are the original potentials given in Eq.~(\ref{pair-pots}) and $\rho_{\rm B}^{(n)}({\bf r})$ is the one body density distribution of B~particles in the molecule centered at~${\bf r}_{\rm C}^{(n)}$, which we model as a Gaussian ring of radius~$r_0$ and width~$w$:
 \begin{equation}\label{eq:rho_B_main_text}
 \rho_{\rm B}^{(n)}({\bf r}) = \rho_0\exp\left(-\frac{(|{\bf r}-{\bf r}_{\rm C}^{(n)}|-r_0)^2}{w^2}\right).
 \end{equation}
The prefactor $\rho_0$ is determined by the normalization condition, $\int\rho_{\rm B}^{(n)}({\bf r})\,d{\bf r}=N$. %(see Eq.~(\ref{eq:rho_B})).
We treat $r_0$, $w$, $g_{\rm AB}$ and~$g_{\rm BB}$, as fitting parameters.
The factors $g_{\rm AB}$ and~$g_{\rm BB}$ describe pair correlation effects, as explained in detail in the SI.
Note that to a good approximation $r_0\approx\ell_0$ and also roughly $w^2\approx2/\beta k_\mathrm{D}$.
Equation~(\ref{eqn:Coarse-grained-PP}) provides a good fit as long as the molecules are not too sparse, so the B~particles in the outer shell behave as a ring rather than as individual arms.
For the parameter values considered here, our data shows this occurs for $N\geq 18$. 

To simplify further, we assume that $\eps_{\rm AA} = 0$, so the architecture of the molecule is specified by $N$, $\eps_{\rm AB}$, $\eps_{\rm BB}$, $R_{\rm AB}$, $R_{\rm BB}$, $k_{\rm D}$ and~$\ell_0$.
We have performed MC simulations for various choices of these parameters to estimate the potential~$V(r)$, and show in Fig.~\ref{fig:2} an example of least squares fitting to the MC data of the effective potential in Eq.~(\ref{eqn:Coarse-grained-PP}).
The fit is very good, suggesting that even with the simplifying assumptions we have taken, the pair potential in Eq.~(\ref{eqn:Coarse-grained-PP}) is representative of the simulation data.

\begin{figure}
\centering
\includegraphics[width=0.98\columnwidth]{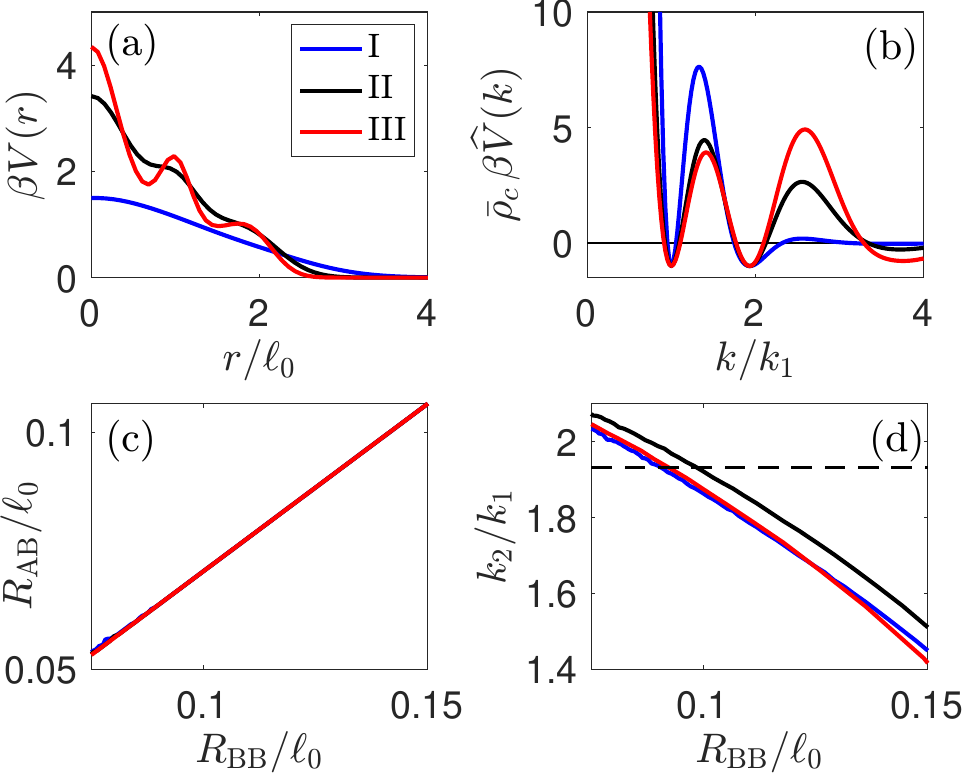}
\caption{Examples of three potentials obtained from the coarse-grained pair potential~(\ref{eqn:Coarse-grained-PP}) for $(w/\ell_0)^2 = \frac{1}{2}$ (I), $\frac{1}{8}$ (II) and $\frac{1}{18}$ (III). We present each potential (a) in real and (b) in Fourier space at parameter values giving QC lengthscales.
Panel~(c) plots the value of $R_{\rm BB}$ as a function of $R_{\rm AB}$ for which these pair potential families have two equal minima, with panel~(d) showing the ratio between the two wavenumbers $k_2/k_1>1$ as a function of $R_{\rm BB}$.
The dashed horizontal line marks the DDQC lengthscale of $2\cos (\pi/12)\approx1.93$.}
\label{fig:3}
\end{figure}
     
With the form of the effective potential determined, we explore the range of possible lengthscales it possesses by performing a parameter scan in the $R_{\rm AB}$-$R_{\rm BB}$ space. 
Recall that this analysis is best done by considering the Fourier transform $\widehat{V}(k)$ \cite{bdl11, ark13, bel14, ark15, Savitz2018, Ratliff2019, Subramanian2021}.
The outcome of this for various values of~$w$ are in Fig.~\ref{fig:3}. 
In these surveys, we use $r_0=\ell_0$, $N=18$ and $g_{\rm BB}\beta\eps_{\rm BB}=1$.
We choose $g_{\rm AB}\beta\eps_{\rm AB}$ such that $V(0)/V(\ell_0)$ is the same across all potentials, to permit us to focus on the impact of the spatial structure of the potential on crystal viability.
The three cases presented are for $(w/\ell_0)^2 = \frac{1}{2}$, $\frac{1}{8}$ and $\frac{1}{18}$.
Fig.~\ref{fig:3}(c) demonstrates that when we require that the two minima of the Fourier transform~$\widehat V(k)$ are equal in value, this is equivalent to $R_{\rm AB}$ being proportional to~$R_{\rm BB}$.
In Fig.~\ref{fig:3}(d), we show the corresponding wavenumber ratios $k_2/k_1$; we see a wide range of ratios are possible, including the value $2\cos(\pi/12)\approx1.93$, which favors dodecagonal QCs (DDQCs).

The results presented here also demonstrate that the value of $w$ affects the value of $\widehat{V}(k)$ (energetic penalty) for wavenumbers $k$ away from the two main lengthscales at $k_1$ and $k_2>k_1$; see Fig.~\ref{fig:3}(b). 
This is known to determine whether crystal structures having lattice vectors at these other wavenumbers are favorable or not \cite{Ratliff2019}. 
Smaller values of $w$, corresponding to a tighter B-particle ring and more prominent spatial structure in $V(r)$, create higher penalties away from the main lengthscales, which has been shown to favor the formation of the patterns associated with these wavevectors~\cite{Ratliff2019}. However, with smaller $w$ we also observe additional (negative) minima in $\widehat{V}(k)$ at higher $k$ values becoming more prominent, which also impact crystal stability and QC formation. 
All of these factors relating to the parameters $w$, $R_{\rm AB}$ and $R_{\rm BB}$ have a bearing on what ultimately determines the stable structures. 
However, there is enough flexibility to permit us to select which lengthscales arise and to control the competition giving rise to specific crystal/QC structures.
%It is this insight that we leverage to explore the crystalline phases the molecules form and tune them to increase or reduce QC viability. 

\begin{figure}
\centering
\includegraphics[width=0.95\columnwidth]{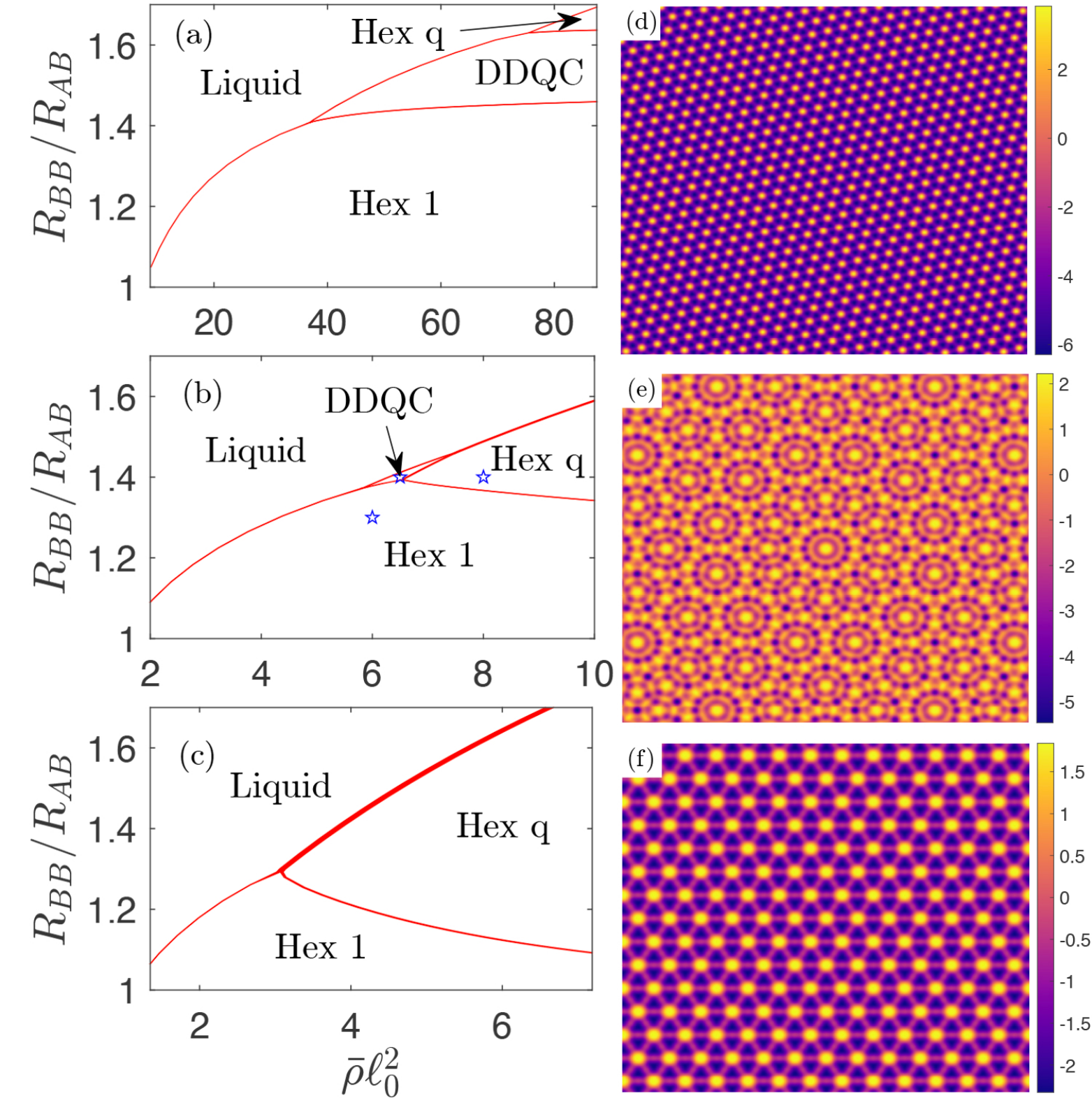}
\caption{Phase diagrams in the average density versus $R_\textrm{BB}/R_\textrm{AB}$ plane, for the three coarse-grained potentials I--III in Fig.~\ref{fig:3}, in panels (a)-(c), respectively. The coexistence regions between neighboring phases are colored red. Logarithms of typical density distributions for each phase in case II appear in panels (d)-(f) with stars in (b) indicating their location in the phase diagram.}
\label{fig:4}
\end{figure}

We investigate the resulting phase diagrams using classical DFT \cite{hansen_mcdonald}. 
With the random-phase-approximation, which is known to be accurate for soft-core systems \cite{bdl11, ark13, bel14, ark15, Savitz2018, Ratliff2019, Subramanian2021, l01, hansen_mcdonald}, this requires us to minimize the grand potential functional
\begin{equation}
\beta\Omega[\rho] = \int\left[\rho (\ln (\rho)-1)+\frac{1}{2}\rho \ (\beta V\circledast \rho)- \beta\mu \rho  \right]\,d{\bf r}, 
\end{equation}
with respect to the density field $\rho({\bf r})$, where $\mu$ is the chemical potential and $\circledast$ denotes a spatial convolution.
 The equilibrium density profiles $\rho({\bf r})$ which minimize $\Omega[\rho]$ satisfy the Euler--Lagrange equation
 \begin{equation}
 \frac{\delta \Omega}{\delta \rho} = \frac{1}{\beta} \ln \rho + V \circledast \rho - \mu = 0\,.
 \end{equation}
We minimize the grand potential across a range of $\mu$ and $R_{\rm AB}$ values.
We fix all other parameters to those used in Fig.~\ref{fig:3} and select the $R_{\rm BB}/\ell_0$ values that result in the lengthscale ratio which supports DDQC formation, $k_2/k_1 = 2\cos(\pi/12)$, i.e., $R_{\rm BB}/\ell_0 = 0.0905,\, 0.0985,\,0.0922$, for cases I-III, respectively.
This results in the three phase diagrams in Fig.~\ref{fig:4}.
For the parameters in case I, we observe a large region of stable DDQCs in the phase diagram.
However, on decreasing $w$, going to cases II and III, we observe the region of stable DDQCs decreases and then disappears, as the competing hexagonal crystal structures (with lattice spacings denoted by 1 and $q$, respectively), become more stable.
Given that going from I to III corresponds to having increasing (real-space) structure in $V(r)$, this result is rather surprising.
This emphasizes again the importance of analyzing such systems in Fourier-space.

%\amr{Will we show a molecular dynamics simulation at these parameter values?}

We have demonstrated here how insight into lengthscale competition may be used to design complex molecules (such as dendrimers or star polymers) in order for self-assembly into a given mesoscopic QC or crystalline structure to occur.
%We have shown how molecular features can be tuned to better favor these.
Strikingly, we have found that systems (case~I) having a {\em less} structured effective potential $V(r)$, have the largest stable region of DDQCs in the phase diagram.
This is due to the fact that when $V(r)$ has this smoother form, density modes at higher $k$ are less favored.
We have analyzed here a 2D system, but the approach naturally extends to 3D, at greater computational expense.
The work presented here constitutes a theoretical framework in which to develop experimentally systems which form quasicrystals, going from molecular architecture to effective interaction potentials and on to phase diagrams.
This whole approach can be applied not just to star polymers and dendrimers, but also to micellar systems and other molecular geometries.
 %Furthermore, one may extend the type of molecule considered from micellar structures to other polymer-based matter such as dendrimers and star polymers using the same methodology, but we expect that the relation between constituent polymers and the emergent lengthscales will differ, owing to the changes in molecular geometry. 

\begin{acknowledgments}
We gratefully acknowledge stimulating discussions with Maria Sammalkorpi. 
This work was supported by the Engineering and Physical Sciences Research Council [grant numbers EP/P015689/1 and EP/P015611/1]; by the Leverhulme Trust [grant number RF-2018-449/9] (AMR); by the London Mathematical Society [Research Reboot grant numbers 42042 and 42043] (PS and DJR); and by the Swiss National Science Foundation [grant number P2FRP2$\_$181453] (AS).
% This work was undertaken on ****, part of the High Performance Computing facilities at ***.
% The data associated with this paper are openly available from the University of **** Data Repository
% (\url{https://doi.org/****})~\cite{Ratliff2026}, as are the programs that generated the data.
For the purpose of open access, the authors have applied a Creative Commons Attribution (CC~BY) license to any Author Accepted Manuscript version arising from this submission.
\end{acknowledgments}

 \pagebreak

% \bibliographystyle{apsrev4-2}
% \bibliography{main.bib}


 %\clearpage
 \onecolumngrid
 \vspace{115mm}

 \pagebreak
 \begin{center}
 {\Large \bf Supplementary information for:\\
Tuning polymer architecture for quasicrystal self-assembly}\\
\vspace{1mm}
{\bf D.~J.~Ratliff, A.~Scacchi, P.~Subramanian, A.~J.~Archer and A.~M.~Rucklidge}
 \end{center}

\setcounter{equation}{0}
\renewcommand{\theequation}{S\arabic{equation}}

%---------------------------------------------------------------%

% \subsection*{ZZZ}

The interaction energy between a pair of molecules, such as the two illustrated in Fig.~\ref{fig:micelle}, is a function of the positions of the various A and B Gaussian particles within each molecule.
The positions of the A particles (i.e., the centers of each dendrimer) are denoted $\{{\bf r}_{\rm C}^{(n)},n=1,2\}$, where the index $n$ denotes which of the two dendrimers being considered.
Tethered to each of the A particles are $N$ B-type particles.
We denote the locations of each of the B particle centers by $\{{\bf r}_i^{(n)}; n=1,2; i=1,\dots,N\}$, where the index $i$ counts over each of the B polymers.
Thus, the total energy of a system containing a pair of dendrimers can be written as
\begin{equation}
\begin{gathered}
E = \phi_{\rm AA}(|{\bf r}_{\rm C}^{(1)}-{\bf r}_{\rm C}^{(2)}|)+\sum_{k,n=1}^2\sum_{j=1}^{N}\phi_{\rm AB}(|{\bf r}_{\rm C}^{(n)}-{\bf r}_j^{(k)}|)%\\
+\frac{1}{2}\sum_{k,n=1}^2\sum_{i,j=1}^{N}\phi_{\rm BB}(|{\bf r}_i^{(n)}-{\bf r}_j^{(k)}|)%\\
+\sum_{n=1}^2\sum_{j=1}^{N}\phi_{\rm AB}^{\rm t}(|{\bf r}_{\rm C}^{(n)}-{\bf r}_j^{(n)}|)\,,
\end{gathered}
\label{eq:int_energy}
\end{equation}
where the pair potentials $\phi_\mathrm{AA}(r)$, $\phi_\mathrm{AB}(r)$ and $\phi_\mathrm{BB}(r)$ are given in Eq.~(\ref{pair-pots}) of the main text and the tether potential $\phi_\mathrm{AB}^\mathrm{t}(r)$ is given in Eq.~(\ref{tether-pot}).

\section{Determining $V(r)$ via Langevin Monte Carlo with Umbrella Sampling}\label{sec:LMC}

To determine the effective coarse-grained pair-potential $V(r)$ between pairs of dendrimers, we use Langevin Monte Carlo (MC) simulations together with \emph{umbrella sampling}.
This uses an external potential to weight the phase space in such a way as to bias the sampling of rare events.
There are various ways to implement this, such as infinite well and harmonic potentials. We use the former, following the approach of Mladek {\it et al.}~\cite{mlk08}, who applied it to dendrimers.
A good introduction to this form of rare events sampling is given in Chapter 6.3 of Ref.~\cite{c87}.
This helps to sample the interactions between the molecules when their separation distance is small, where such configurations are unlikely, due to the repulsive forces.
The key step is to create a sequence of windows (each indexed with the integer $p$) using the infinite well external potentials of the form
\begin{equation}
U_i({\bf r}) = \begin{cases}
0\,, \quad {\bf r} \in [{\bf r}^{(p)}_{\rm L},{\bf r}^{(p)}_{\rm U}],\\
\infty, \quad {\rm otherwise,}
\end{cases}
\end{equation}
which is applied to the centers of mass of each of the dendrimers, i.e., ${\bf r}={\bf r}_\mathrm{COM}^{(n)}$, $n=1,2$, where ${\bf r}_\mathrm{COM}^{(n)}$ is the center of mass of dendrimer $n$. 
%Note that ${\bf r}_\mathrm{COM}^{(n)}$ is not {\em exactly} the same as ${\bf r}_\mathrm{C}^{(n)}$, but the distinction is sufficiently small that for present purposes, the small difference can be neglected.
We then simulate the evolution of the pair of molecules using Langevin MC with this external potential present.
Any MC move which leads to a molecule's center of mass lying outside of the `window' is rejected.
We sample many such windows, with boundaries which overlap slightly, over the domain and generate a radial distribution function for each window, $g_p(r)$, where $r$ is the distance between the centers of the two molecules. The final step is to stitch these together.
Since the radial distribution function is correct up to a multiplicative constant within each window, we make sure that each portion of the radial distribution function agrees with its neighbor within the overlap region of the windows, i.e.:
\begin{equation}
g_p({r}^{(p)}_{\rm U}) = g_{p-1}({r}^{(p-1)}_{\rm L})\,,    
\end{equation}
which is achieved via the mapping
\begin{equation}
g_p \mapsto \frac{g_{p-1}({r}^{(p-1)}_{\rm L})}{g_p({r}^{(p)}_{\rm U})}g_p\,.
\end{equation}
Finally, we scale everything so that the right-most window $g_1$ satisfies $g_1(r^{(1)}_{\rm U}) = 1$. This generates a unique distribution function with improved sampling at shorter pair separation distances $r$.

For our MC moves, we use the standard Metropolis algorithm. To implement this, one chooses randomly one of the particles within one of the dendrimers.
A uniform random movement of the selected particle is then applied. One then checks if this move is accepted based on the ratio between the probability of the states before and after the move, $\alpha = p_{\rm new}/p_{\rm old}$.
This is then compared to a random number $\gamma$ obtained uniformly from the range [0,1]. If $\alpha>\gamma$ then the move is accepted, else it is rejected and the state prior to the move is retained.
This always accepts moves that decrease the energy $E$ in Eq.~\eqref{eq:int_energy}, since $\alpha>1$ means that $p_{\rm new}>p_{\rm old}$, but it can also sample moves that increase $E$ with a probability dependent on the increase in energy.
To calculate the probabilities, we recall that they are proportional to the Boltzmann factor, $p \propto \exp(-\beta E)$. 
From this we can then see that the transition probability $\alpha$ can be written as
\begin{equation}\label{alpha-E}
\alpha = \exp\left[-\beta(E_{\rm new}-E_{\rm old})\right].
\end{equation}
In the \emph{low density limit}, the effective molecule-molecule interaction potential $V(r)$ can be obtained from the radial distribution function $g(r)$ using the relation~\cite{c87}
\begin{equation}
V(r) = -k_{\rm B}T\ln g(r).
\end{equation}
A typical result from applying this approach is displayed in Fig.~\ref{fig:2}.

To check the results obtained from the umbrella sampling Langevin MC simulations, we also used standard Brownian dynamics (BD) simulations of the pair of molecules to calculate $g(r)$ (not displayed).
The BD simulations are computationally much more expensive, but allow to obtain the same result, thereby confirming the reliability of our MC simulations scheme.
The BD simulations of the pair of molecules were performed in a (large) square box of size $30 \times 30\>r_0^2$.
We carried out 50 independent simulations with integration time step $dt=2\times 10^{-3}$ for $10^9$ steps, while setting the bare diffusion of each block to $D=1$.
The data was sampled every $10^3$ steps for the analysis.

%In Fig.~(X) we compare $V(r)$ obtained from Brownian Dynamics simulations and Langevin Monte Carlo, for two sets of parameters (see figure caption). The very good agreement between the two curves provide a solid validation of the Langevin Monte Carlo simulation method for the coarse-graining of complex molecular superstructures.

\section{Effective interaction potential}

To coarse-grain, we seek to define an effective interaction potential~\cite{l01} between a pair of molecules, defined as a function of the position coordinates of each of the central A `hubs'.
Thus, we must perform an ensemble average over all possible configurations of all the B~particles in each of the molecules, for fixed positions of the A~particles.
To do this, we rewrite Eq.~\eqref{eq:int_energy} as:
\begin{eqnarray}
%\begin{gathered}
E &=& \phi_{\rm AA}(|{\bf r}_{\rm C}^{(1)}-{\bf r}_{\rm C}^{(2)}|)%\\
+\sum_{k,n=1}^2\sum_{j=1}^{N}\int\int\phi_{\rm AB}(|{\bf r}-{\bf r}'|)\delta({\bf r}-{\bf r}_j^{(k)})\delta({\bf r}'-{\bf r}_{\rm C}^{(n)})d{\bf r}d{\bf r}'\nonumber \\
&\,&{}+\frac{1}{2}\sum_{k,n=1}^2\sum_{i,j=1}^{N}\int\int\phi_{\rm BB}(|{\bf r}-{\bf r}'|)\delta({\bf r}-{\bf r}_i^{(n)})\delta({\bf r}'-{\bf r}_j^{(k)})d{\bf r}d{\bf r}'\nonumber\\
&\,&{}+\sum_{n=1}^2\sum_{j=1}^{N}\int\int\phi_{\rm AB}^{\rm t}(|{\bf r}-{\bf r}'|)\delta({\bf r}-{\bf r}_j^{(n)})\delta({\bf r}'-{\bf r}_{\rm C}^{(n)})d{\bf r}d{\bf r}'\,,
%\end{gathered}
\label{eq:int_energy_with_deltas}
\end{eqnarray}
where $\delta({\bf r})$ denotes a 2D Dirac delta function.
The effective interaction potential between the pair of molecules is then defined as the ensemble average of the energy, with the centers of the two fixed at the specific points ${\bf r}_{\rm C}^{(1)}$ and ${\bf r}_{\rm C}^{(2)}$,
\begin{align} % requires amsmath; align* for no eq. number
   V({\bf r}_{\rm C}^{(1)}-{\bf r}_{\rm C}^{(2)})\equiv\left\langle E({\bf r}_{\rm C}^{(1)}-{\bf r}_{\rm C}^{(2)})\right\rangle\,,
\end{align}
where $\left\langle\cdots\right\rangle$ denotes the ensemble average.
We should emphasize that this averaging is over all possible positions of the B particles; the two A particles are taken to be at either ${\bf r}_{\rm C}^{(1)}$ or ${\bf r}_{\rm C}^{(2)}$.
Thus, from Eq.~\eqref{eq:int_energy_with_deltas}, we obtain
\begin{eqnarray}
V({\bf r}_{\rm C}^{(1)}-{\bf r}_{\rm C}^{(2)}) &=& \phi_{\rm AA}(|{\bf r}_{\rm C}^{(1)}-{\bf r}_{\rm C}^{(2)}|)
+\sum_{k,n=1}^2\int\int\phi_{\rm AB}(|{\bf r}-{\bf r}'|)\left\langle \delta({\bf r}'-{\bf r}_{\rm C}^{(n)})\sum_{j=1}^{N}\delta({\bf r}-{\bf r}_j^{(k)})\right\rangle d{\bf r}d{\bf r}'\nonumber\\
&\,&+{}\frac{1}{2}\sum_{k,n=1}^2\int\int\phi_{\rm BB}(|{\bf r}-{\bf r}'|)\left\langle\sum_{i=1}^{N}\delta({\bf r}-{\bf r}_i^{(n)})\sum_{j=1}^{N}\delta({\bf r}'-{\bf r}_j^{(k)})\right\rangle d{\bf r}d{\bf r}'\nonumber\\
&\,&+{}\sum_{n=1}^2\int\int\phi_{\rm AB}^{\rm t}(|{\bf r}-{\bf r}'|)\left\langle\delta({\bf r}'-{\bf r}_{\rm C}^{(n)})\sum_{j=1}^{N}\delta({\bf r}-{\bf r}_j^{(n)})\right\rangle d{\bf r}d{\bf r}'\,.
\label{eq:int_energy_with_deltas_average}
\end{eqnarray}
Note that the term in the second line involves the two-body distribution function \cite{hansen_mcdonald}
\begin{align}
\rho_{\rm BB}^{(nk)}({\bf r},{\bf r}')\equiv 
\left\langle\sum_{i=1}^{N}\delta({\bf r}-{\bf r}_i^{(n)})\sum_{j=1}^{N}\delta({\bf r}'-{\bf r}_j^{(k)})\right\rangle\,.
\end{align}
Below, we relate this to the (one body) density distribution of B particles within each molecule,
\begin{align}
\rho_{\rm B}^{(n)}({\bf r})\equiv \left\langle\sum_{j=1}^{N}\delta({\bf r}-{\bf r}_j^{(n)})\right\rangle\,,
\label{eq:rho1_B}
\end{align}
but for now we write in terms of the two-body function.
Note also that $\rho_{\rm BB}^{(11)}$ and $\rho_{\rm BB}^{(22)}$ are the intra-molecule B-B pair correlation functions, while $\rho_{\rm BB}^{(12)}$ is the corresponding cross-correlation function between the positions of the B particles in the two different molecules.
Similarly, the second term on the right hand side of the first line of Eq.~\eqref{eq:int_energy_with_deltas_average} and also the term in the third line involve the quantity
\begin{align}
\rho_{\rm AB}^{(nk)}({\bf r},{\bf r}')\equiv 
\left\langle \delta({\bf r}-{\bf r}_{\rm C}^{(n)})\sum_{j=1}^{N}\delta({\bf r}'-{\bf r}_j^{(k)})\right\rangle\,,
\label{eq:rhoAB}
\end{align}
which is an A-B two-body distribution function. We should emphasize again that this is somewhat unusual because we have constrained the two A particles to be at one of the two locations ${\bf r}_{\rm C}^{(1)}$ or~${\bf r}_{\rm C}^{(2)}$.
Using the two-body distribution functions defined above, we can re-write Eq.~\eqref{eq:int_energy_with_deltas_average} as:
\begin{eqnarray}
V({\bf r}_{\rm C}^{(1)}-{\bf r}_{\rm C}^{(2)}) &=& \phi_{\rm AA}(|{\bf r}_{\rm C}^{(1)}-{\bf r}_{\rm C}^{(2)}|)
+\sum_{k,n=1}^2\int\int\phi_{\rm AB}(|{\bf r}-{\bf r}'|)\rho_{\rm AB}^{(nk)}({\bf r},{\bf r}') d{\bf r}d{\bf r}'\nonumber\\
&\,&{}+\frac{1}{2}\sum_{k,n=1}^2\int\int\phi_{\rm BB}(|{\bf r}-{\bf r}'|)\rho_{\rm BB}^{(nk)}({\bf r},{\bf r}') d{\bf r}d{\bf r}'\nonumber\\
&\,&{}+\sum_{n=1}^2\int\int\phi_{\rm AB}^{\rm t}(|{\bf r}-{\bf r}'|)\rho_{\rm AB}^{(nn)}({\bf r},{\bf r}') d{\bf r}d{\bf r}'\,.
\label{eq:int_energy_with_rhos}
\end{eqnarray}
%Don't need next equation
%\begin{eqnarray}
%V({\bf r}_{\rm C}^{(1)}-{\bf r}_{\rm C}^{(2)}) &=& N\phi_{\rm AA}(|{\bf r}_{\rm C}^{(1)}-{\bf r}_{\rm C}^{(2)}|)
%+N\sum_{k,n=1}^2\int\phi_{\rm AB}(|{\bf r}_{\rm C}^{(n)}-{\bf r}|)\rho_B^{(k)}({\bf r};{\bf r}_{\rm C}^{(k)}) d{\bf r}\nonumber\\
%&\,&{}+\frac{1}{2}\sum_{k,n=1}^2\int\int\phi_{\rm BB}(|{\bf r}-{\bf r}'|)\rho_{\rm BB}({\bf r},{\bf r}';{\bf r}_{\rm C}^{(n)}-{\bf r}_{\rm C}^{(k)}) d{\bf r}d{\bf r}'\nonumber\\
%&\,&{}+\frac{1}{N}\sum_{n=1}^2\int\int\phi_{\rm AB}^{\rm t}(|{\bf r}_{\rm C}^{(n)}-{\bf r}|)\rho_{\rm AB}({\bf r},{\bf r}';{\bf r}_{\rm C}^{(n)}-{\bf r}_{\rm C}^{(k)}) d{\bf r}d{\bf r}'\,.
%\label{eq:int_energy_with_rhos}
%\end{eqnarray}
Consider now the terms in the last line: these are just the contribution from the tether potential between the A-particles in the `hub' of each molecule with the surrounding B-particles.
In other words, this is an intra-molecule contribution to the energy that does not depend directly on the distance between the pair of molecules ${\bf r}_{\rm C}^{(1)}-{\bf r}_{\rm C}^{(2)}$.
Of course, the relative positions are involved in determining the shape of the B-particle density distribution $\rho_{\rm B}^{(n)}({\bf r})$ (through the ensemble average) and therefore indirectly to the pair interaction $V$.
Thus, we should expect this whole term to only weakly depend on the distance between the molecules $({\bf r}_{\rm C}^{(1)}-{\bf r}_{\rm C}^{(2)})$, and we assume it to be well-approximated by a constant, which therefore can be neglected.

We introduced above in Eq.~\eqref{eq:rho1_B} the one-body distribution function of the B particles, $\rho_{\rm B}^{(n)}({\bf r})$.
We have not quite got to the stage where we can introduce this quantity into our expression for~$V$.
However, it will help the reader if we discuss now the approximation we make for this quantity, to give a better idea of where we are headed.
Here, we assume the following Gaussian form for the density distribution of the B particles in each molecule
\begin{equation}
\rho_{\rm B}^{(n)}({\bf r})=\rho_{\rm B}^{(n)}(|{\bf r}-{\bf r}_{\rm C}^{(n)}|) = \rho_0\exp \left(-\frac{(|{\bf r}-{\bf r}_{\rm C}^{(n)}|-r_0)^2}{w^2}\right),
\label{eq:rho_B}
\end{equation}
where $\rho_0$ is a normalization constant determined by the requirement that
\begin{equation}
\label{eq:normalisation}
    \int \rho_{\rm B}^{(n)}(|{\bf r}-{\bf r}_{\rm C}^{(n)}|)d{\bf r}=N\,,
\end{equation}
where $N$ is the number of B particles in each molecule. The other two parameters in Eq.~\eqref{eq:rho_B} are~$r_0$, the average distance of each B particle from its respective `hub', and~$w$, the average width of the thermally driven wanderings away from the average position. Here, both $r_0$ and $w$ are treated as fit parameters, which we obtain from the simulations.
With the assumption in Eqs.~\eqref{eq:rho_B}, performing the integration in \eqref{eq:normalisation}, we obtain
\begin{equation}
    \rho_0 \pi w \left( \sqrt{\pi}r_0\left[1+{\rm erf}\left(\frac{r_0}{w}\right)\right]+w\exp\left[-\left(\frac{r_0}{w}\right)^2\right]\right)=N,
\end{equation}
where ${\rm erf}(x)$ is the standard error function.
Moreover, in the case relevant here, where the radius of the B-particle ring is larger than its width, we have $r_0>w$, so that ${\rm erf}({r_0}/{w})\approx 1$ to a good approximation and also $\exp(-({r_0}/{w})^2)\approx 0$, giving
\begin{equation}
    \rho_0\approx\frac{N}{2\pi^{3/2} w r_0}.
\end{equation}

Returning now to Eq.~\eqref{eq:int_energy_with_rhos}, we first consider the terms in the second line. A standard way of re-writing a two-body density distribution function in terms of the one-body density profiles is \cite{hansen_mcdonald}
\begin{equation}
    \rho_{\rm BB}^{(nk)}({\bf r},{\bf r}')=
    \rho_{\rm B}^{(n)}({\bf r})
    \rho_{\rm B}^{(k)}({\bf r}')
    g_{\rm BB}({\bf r},{\bf r}'),
\end{equation}
where $g_{\rm BB}$ is an inhomogeneous fluid (of B particles) pair correlation function.
A common approximation, often referred to as the random-phase-approximation (RPA) \cite{hansen_mcdonald}, is to neglect the correlations and assume $g_{\rm BB}(r)\approx 1$ for all~$r$.
However, in view of the way the B particles can avoid overlaps with each other (see for example the snapshots in Fig.~\ref{fig:2}), an assumption that there are no pair correlations is not appropriate.
Here, we assume that $g_{\rm BB}(r)=g_{\rm BB}<1$, is a constant. In other words, that the B-particles are able (on average) to avoid each other, but that there are no spatial correlations in these avoidances. Again, we treat the constant $g_{\rm BB}$ as a fit parameter.
In the large $N\to\infty$ limit we expect to recover $g_{\rm BB}=1$ to be a good approximation.
In contrast, in the limit when $N$ is small, then $g_{\rm BB}\approx0$, since most of the time the B-particles can avoid each other.

Similarly, for the A--B pair correlation function \eqref{eq:rhoAB}, we write it as
\begin{equation}
    \rho_{\rm AB}^{(nk)}({\bf r},{\bf r}') = \delta({\bf r}-{\bf r}_{\rm C}^{(n)})\rho_{\rm B}^{(k)}({\bf r}')g_{\rm AB}
\end{equation}
where similarly to the B--B case, we assume that the A--B pair correlation function $g_{\rm AB}(r)=g_{\rm AB}<1$, is a constant. 

Putting all these approximations into Eq.~\eqref{eq:int_energy_with_rhos}, we obtain
\begin{eqnarray}
V({\bf r}_{\rm C}^{(1)}-{\bf r}_{\rm C}^{(2)}) &=& \phi_{\rm AA}(|{\bf r}_{\rm C}^{(1)}-{\bf r}_{\rm C}^{(2)}|)\nonumber\\
&\,&{}+\int\phi_{\rm AB}(|{\bf r}_{\rm C}^{(1)}-{\bf r}|)g_{\rm AB}\rho_{\rm B}^{(2)}({\bf r}) d{\bf r}
+\int\phi_{\rm AB}(|{\bf r}_{\rm C}^{(2)}-{\bf r}|)g_{\rm AB}\rho_{\rm B}^{(1)}({\bf r}) d{\bf r}\nonumber\\
&\,&{}+\frac{1}{2}\sum_{k,n=1}^2\int\int\phi_{\rm BB}(|{\bf r}-{\bf r}'|)\rho_{\rm B}^{(n)}({\bf r})
    \rho_{\rm B}^{(k)}({\bf r}')
    g_{\rm BB} d{\bf r}d{\bf r}'.
\label{eq:int_energy_simp}
\end{eqnarray}
Note that by symmetry, because of our assumption that the distributions of B~particles around each ring are the same, the two terms in the second line are equal.
To summarize, we now have an expression for the effective interaction between a pair of molecules that now only depends on the four fit parameters $r_0$, $w$, $g_{\rm BB}$ and~$g_{\rm AB}$.
Written in this way, $V$~depends only on the distance between the two molecules.
Here, the convolution integrals in Eq.~\eqref{eq:int_energy_simp} are evaluated numerically in MATLAB on a 2D grid using fast Fourier transforms.

 %\bibliography{main.bib}		
 \end{document}